\def\gtorder{\mathrel{\raise.3ex\hbox{$>$}\mkEern-14mu\lower0.6ex\hbox{$\sim$}}}
\def\ltorder{\mathrel{\raise.3ex\hbox{$<$}\mkern-14mu\lower0.6ex\hbox{$\sim$}}}
\def\etal{et al.~}
\def\eg{e.g.~}

\def\kms{$\rm km\,s^{-1}$}
\def\asec{^{\prime\prime}}
\def\Msun{{\rm M}_{\odot}}

\def\figureone{
 \vskip 0.3cm
 \begin{figure*}[t]
 \figurenum{1}
 \centerline{\psfig{file=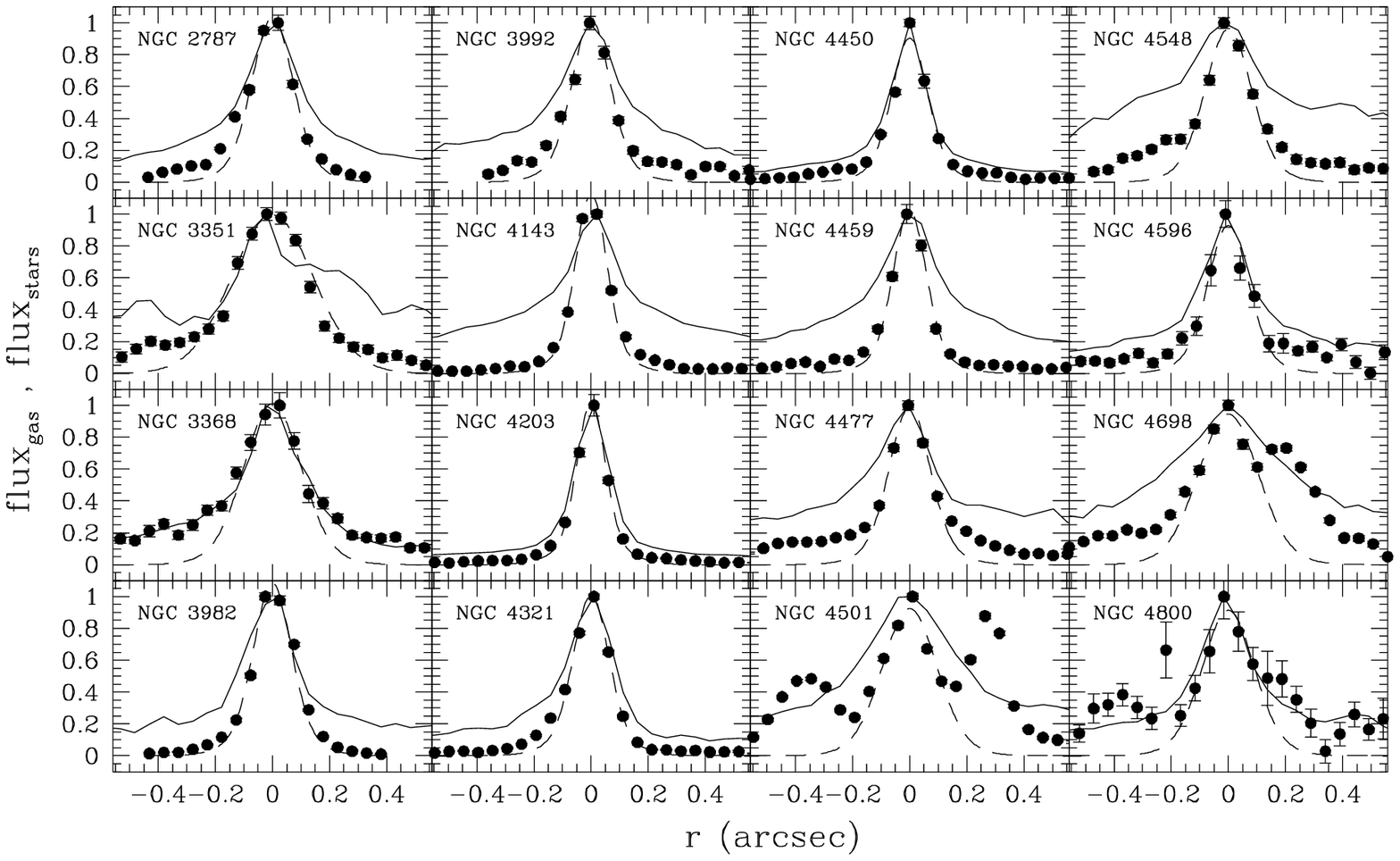,width=18.5cm,angle=0}}
 \figcaption[fig1.ps]{
  Observed emission-line flux profiles along the direction of the STIS
  slit.  In each panel the {\it filled symbols} represent the observed
  [\ion{N}{2}] fluxes at each CCD row along the slit (corresponding to
  a synthesized aperture of $0\farcs05 \times 0\farcs2$), while the
  {\it dashed line} shows the fit to the central five points as
  described in \S \ref{subsec:UppLim_TheDiskModelling}. For comparison
  the {\it solid line} represents the stellar continuum profile,
  extracted from the region free of emission lines between 6536 and
  6542 \AA.  Both emission-line and continuum flux profiles have been
  rescaled so that their peak values equal unity.
 \label{fig:UppLim_FluxProfilesAndFitAndContinuum}}
 \end{figure*}
 \vskip 0.3cm
}
\def\figuretwo{
 \vskip 0.3cm
 \figurenum{2}
 \psfig{file=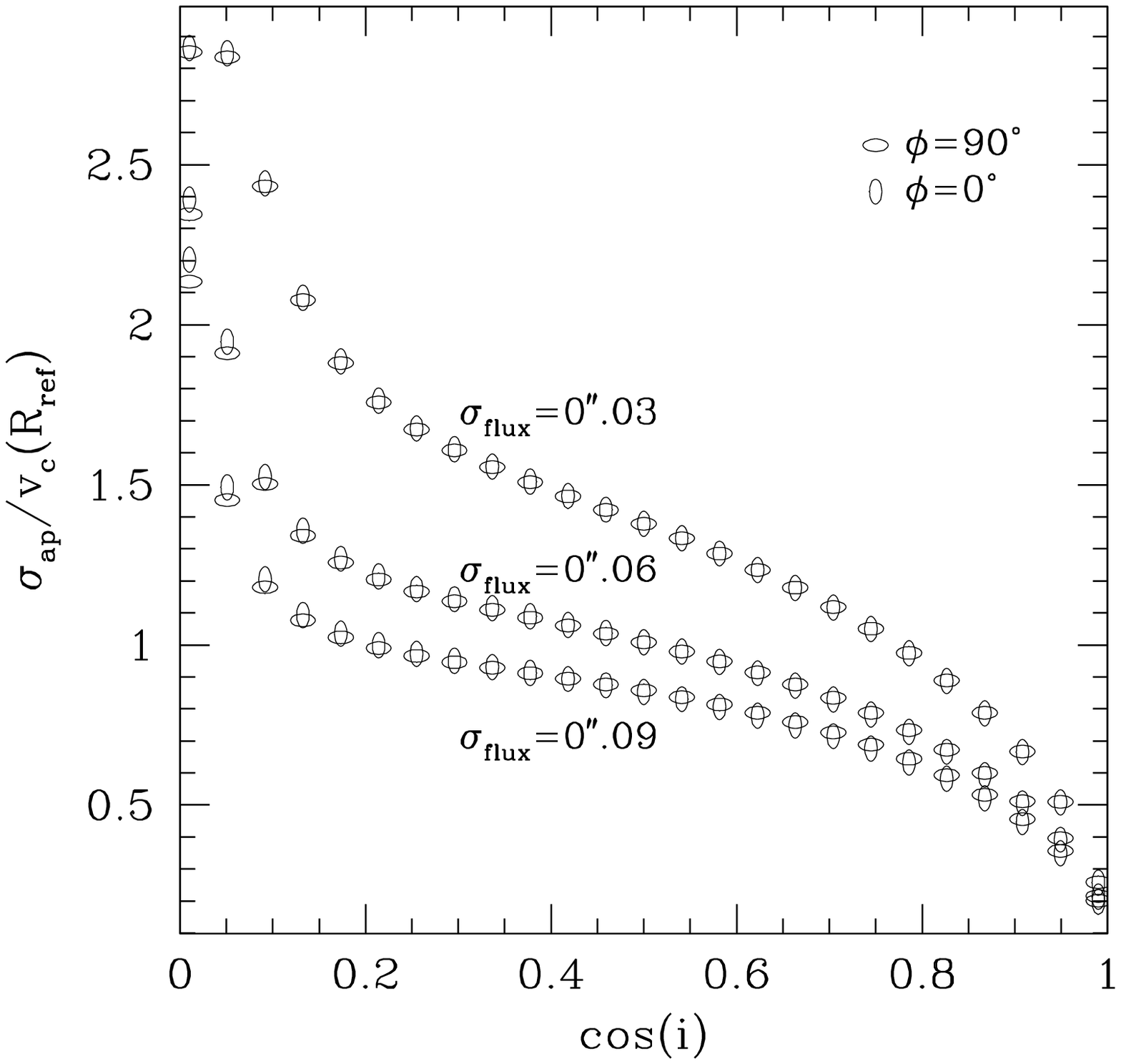,width=8.5cm,angle=0}
 \figcaption[fig2.ps]{
  Predicted ratio of the gas central velocity dispersion to the
  circular velocity at the reference radius, as obtained by the
  Keplerian disk modeling (\S \ref{subsec:UppLim_TheDiskModelling}) for
  different disk inclinations, position angles $\phi$, and intrinsic
  flux distributions.
 \label{fig:UppLim_SigOverVcExample}}
 \vskip 0.3cm
}
\def\figurethree{
 \vskip 0.3cm
 \figurenum{3}
 \psfig{file=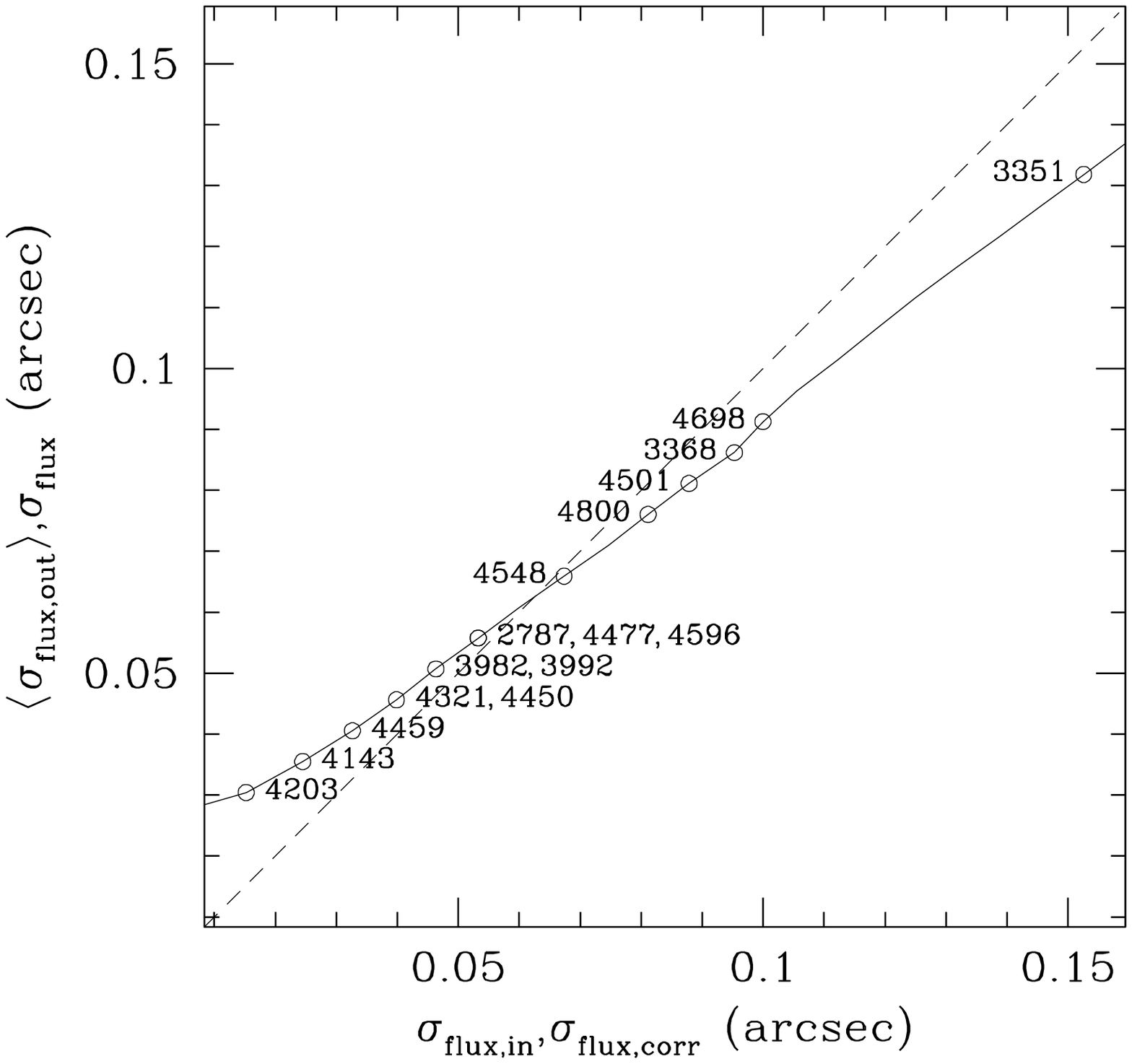,width=8.5cm,angle=0}
 \figcaption[fig3.ps]{
  Correction function for the derived flux concentrations.  The {\it
  solid line\/} relates each intrinsic $\sigma_{flux,in}$ with the
  median $\langle\sigma_{flux,out}\rangle$ of all the
  $\sigma_{flux,out}$ values that well matched the flux profiles
  obtained from each $\sigma_{flux,in}$, as described in
  \S\ref{subsec:UppLim_TheDiskModelling}.  Correspondingly, for each
  sample galaxy, the {\it open circles\/} relate each of the
  $\sigma_{flux}$ obtained by matching the galaxy flux profiles of
  Fig. \ref{fig:UppLim_FluxProfilesAndFitAndContinuum}, with the
  intrinsic flux concentration $\sigma_{flux,corr}$ that need to be
  input into the models. The identity transformation ({\it dashed
  line\/}) is shown for comparison.
 \label{fig:UppLim_SigFluxGuessCorrection}}
 \vskip 0.3cm
}
\def\figurefour{
 \vskip 0.3cm 
 \figurenum{4}
 \psfig{file=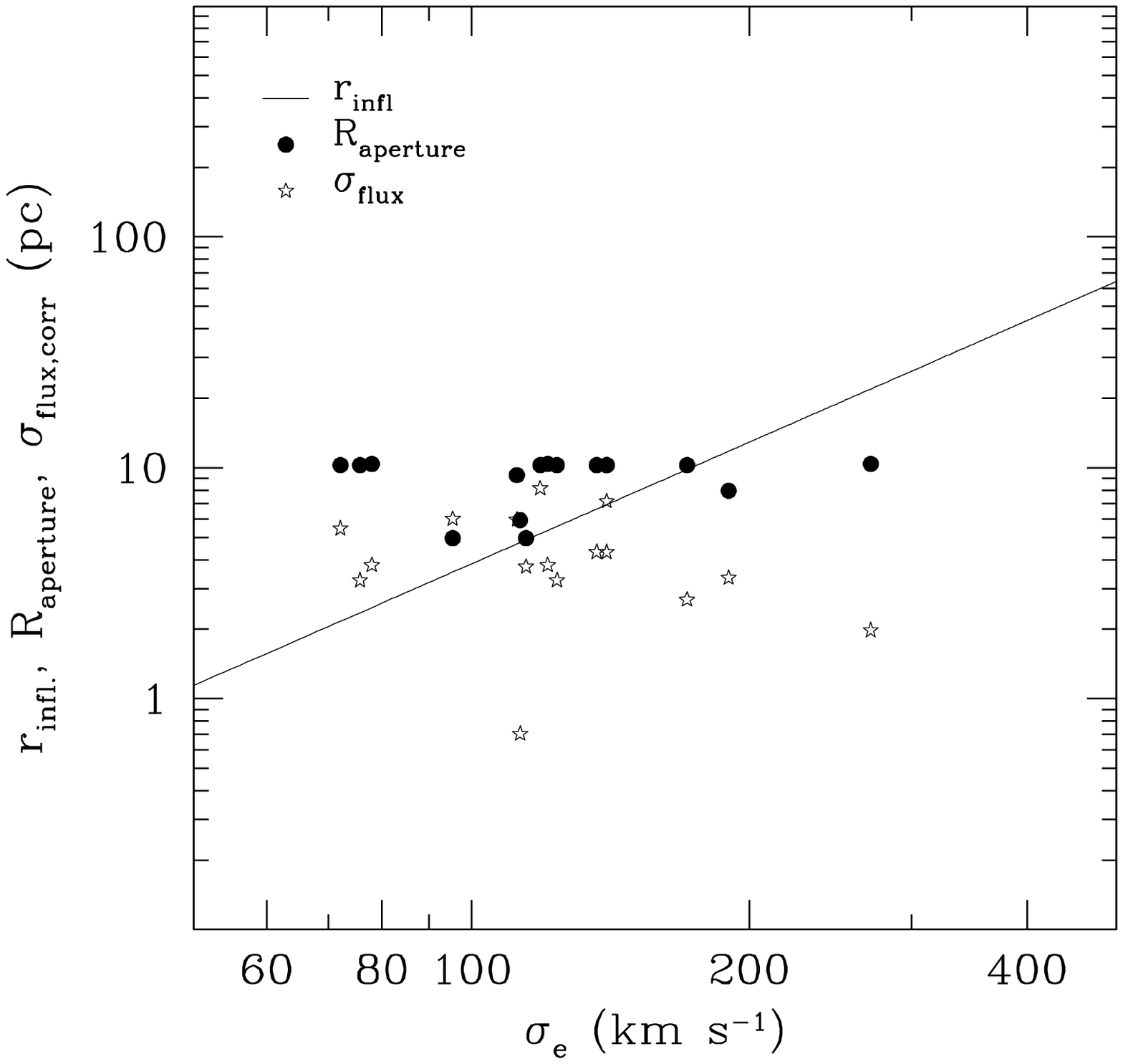,width=8.5cm,angle=0}
 \figcaption[fig4.ps]{
  Physical dimension of the region subtended by our central $0\farcs25
 \times 0\farcs2$ aperture ({\it filled circles\/}) and flux
  distribution extent of our sample galaxies ({\it open stars\/})
  compared to the ``sphere of influence'' radius of SMBHs consistent
  with the $M_{\rm BH}-\sigma_{\star}$ relation ({\it solid line\/}).
 \label{fig:UppLim_Rinfluence}}
 \vskip 0.3cm
}
\def\figurefive{
 \vskip 0.3cm
 \figurenum{5}
 \psfig{file=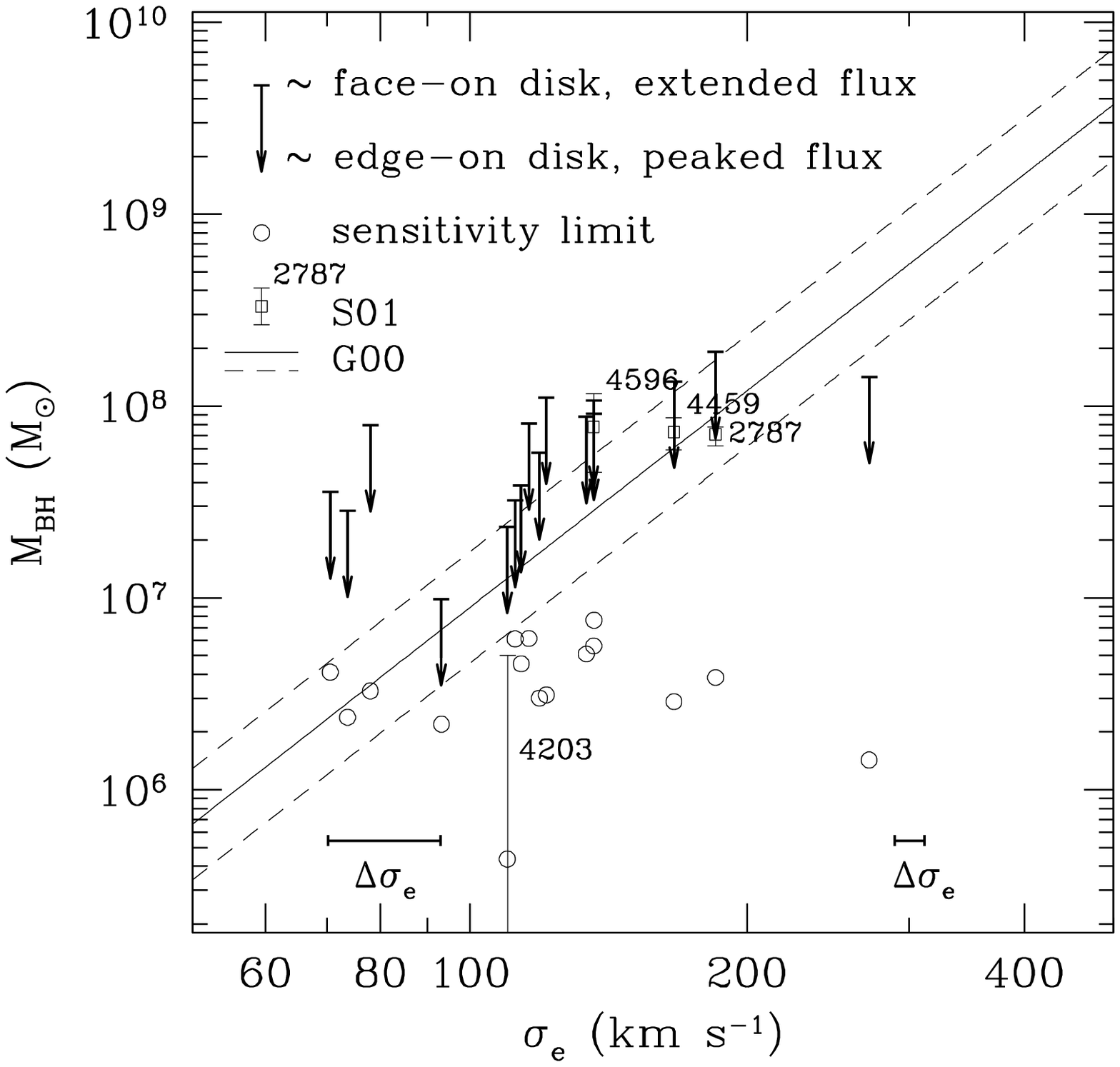,width=8.5cm,angle=0}
 \figcaption[fig5.ps]{
  $M_{\rm BH}$ upper limits in the $M_{\rm BH}-\sigma_{\star}$
  relation.  The $+1\sigma$ upper limits on $M_{\rm BH}$ ({\it thick
  downward arrows \/}) obtained by the Keplerian disk modeling (\S
  \ref{subsec:UppLim_TheDiskModelling}) are compared with the range of
  black-hole masses expected as a function of $\sigma_{\star}$ from the
  $M_{\rm BH}-\sigma_{\star}$ relation and its absolute scatter in
  $M_{\rm BH}$ ({\it solid \/} and {\it dashed lines \/}, from Gebhardt
  \etal 2000a).  The {\it open squares \/} show the $M_{\rm BH}$
  measurements obtained from extended kinematics in four of our sample
  galaxies by S01, while the {\it open circles\/} represent the
  sensitivity limit of our experiment for each sample galaxy. The {\it
  thick horizontal error bars \/} indicate how the mean error on
  $\sigma_{\star}$ for our sample would appear in the right and left
  hand of the plot.
 \label{fig:UppLim_UppLimKeplerian}}
 \vskip 0.3cm
}
\def\figuresix{
 \vskip 0.3cm
 \begin{figure*}[t]
 \figurenum{6}
 \centerline{\psfig{file=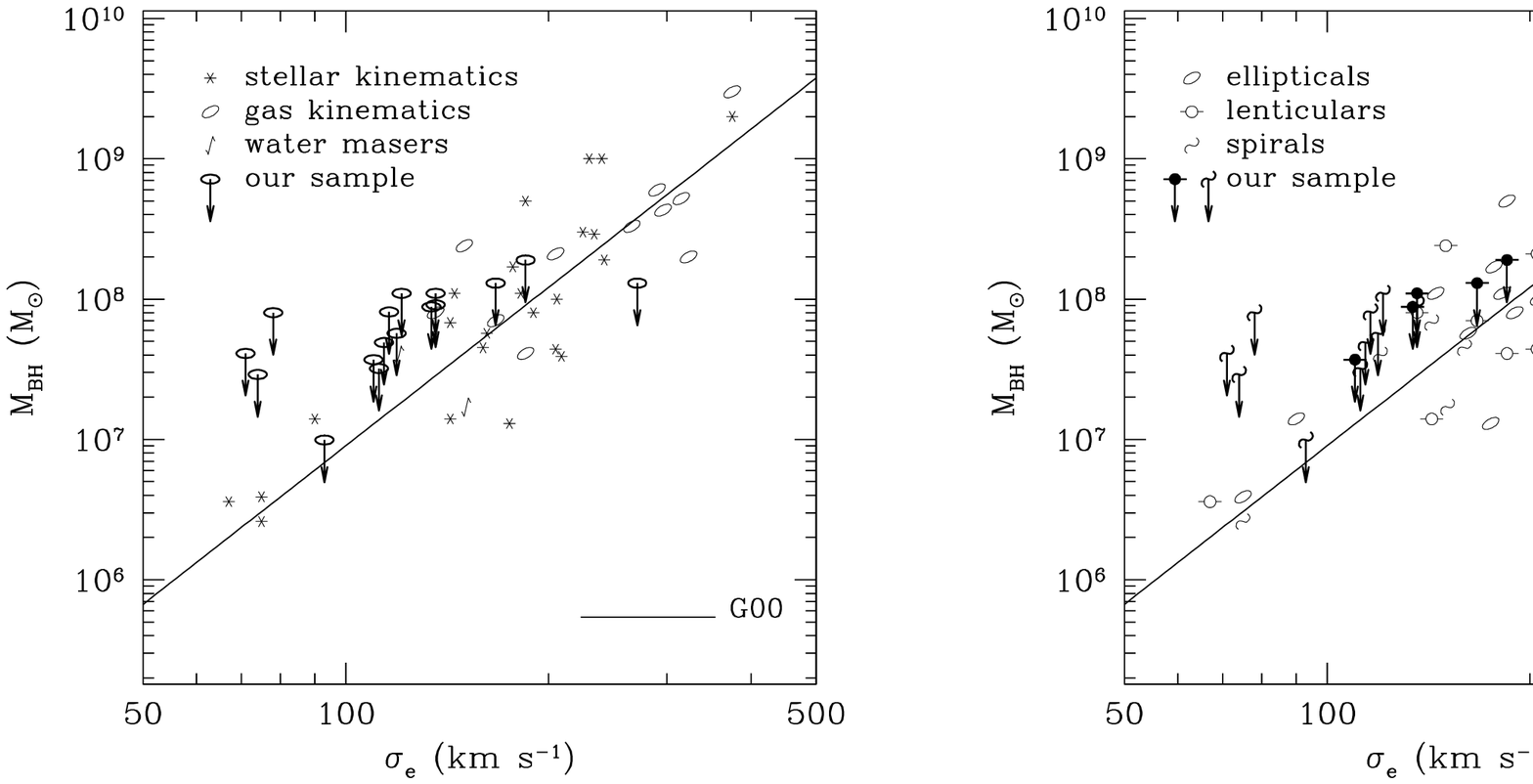,width=19.5cm,angle=0}}
 \figcaption[fig6.ps]{
  $M_{\rm BH}$ upper limits versus $M_{\rm BH}$ determinations. In both
  panels the $M_{\rm BH}$ upper limits for our sample galaxies ({\it
  symbols with downward arrows\/}) are compared with the $M_{\rm BH}$
  determinations that were deemed accurate in the recent compilation of
  Kormendy \& Gebhardt (2001) ({\it symbols without arrows\/}).  To be
  consistent with this compilation, for our sample objects with
  measured surface brightness fluctuation distance moduli (Tonry \etal
  2001, for NGC~2787, NGC~3368, NGC~4143, NGC~4203, NGC~4459, and
  NGC~4548) the $M_{\rm BH}$ upper limits have been rescaled assuming
  these distances.  In the {\it left panel\/} the different symbols
  stand for the different techniques used to measure the $M_{\rm BH}$,
  while in the {\it right panel\/} they represent the different types
  of host galaxies. The {\it solid lines} show the $M_{\rm
  BH}-\sigma_{\star}$ relation as compiled by Gebhardt \etal (2000a).
 \label{fig:UppLim_UppLimKeplerian_G00_Mbhs}}
 \end{figure*}
 \vskip 0.3cm
}

\documentclass[preprint]{aastex}
\usepackage{emulateapj5}
\usepackage{apjfonts}
\usepackage{natbib}

\input psfig.tex

\slugcomment{To appear in {\it The Astrophysical Journal}.}
\lefthead{SARZI ET AL.}
\righthead{BLACK HOLE MASS LIMITS}

\begin{document}

\title{Limits on the Mass of the Central Black Hole in Sixteen Nearby 
Bulges\footnotemark[1]}

\footnotetext[1]{Based on observations made with the {\it Hubble Space
Telescope}, which is operated by AURA, Inc., under NASA contract NAS5-26555.}

\author{
Marc Sarzi\altaffilmark{2,3}, 
Hans-Walter Rix\altaffilmark{2}, 
Joseph C. Shields\altaffilmark{4}, 
Daniel H. McIntosh\altaffilmark{5}, 
Luis C. Ho\altaffilmark{6}, 
Gregory Rudnick\altaffilmark{2,7}, 
Alexei V. Filippenko\altaffilmark{8}, 
Wallace L. W. Sargent\altaffilmark{9}, and 
Aaron J. Barth\altaffilmark{10}
}

\altaffiltext{2}{Max-Planck-Institut f{\"u}r Astronomie, K{\"o}nigstuhl 17, 
  Heidelberg, D-69117, Germany; sarzi, rix, grudnick@mpia-hd.mpg.de.}

\altaffiltext{3}{Dipartimento di Astronomia, Universit\`a di Padova,
  Vicolo dell'Osservatorio 5, I-35122 Padova, Italy;
  sarzi@pd.astro.it.}

\altaffiltext{4}{Physics \& Astronomy Department, Ohio University,
  Athens, OH 45701; shields@phy.ohiou.edu.}

\altaffiltext{5}{Astronomy Department, University of Massachusetts, Amherst, 
  MA 01003; dmac@hamerkop.astro.umass.edu}

\altaffiltext{6}{The Observatories of the Carnegie Institution of
  Washigton, 813 Santa Barbara St., Pasadena, CA 91101-1292;
  lho@ociw.edu.}

\altaffiltext{7}{Steward Observatory, University of Arizona, Tucson,
  AZ 85721; grudnick@as.arizona.edu.}

\altaffiltext{8}{Astronomy Department, University of California,
  Berkeley, CA 94720-3411; alex@astro.berkeley.edu.}

\altaffiltext{9}{Palomar Observatory, Caltech 105-24, Pasadena, CA
  91125; wws@astro.caltech.edu.}

\altaffiltext{10}{Harvard-Smithsonian Center for Astrophysics, 60 
  Garden Street, Cambridge, MA 02138; abarth@cfa.harvard.edu.}

\setcounter{footnote}{10}

\begin{abstract}
  We report upper limits on the masses of black holes that can be
  present in the centers of sixteen nearby galaxy bulges.  These
  limits $M_{\rm BH}^{lim}$ for our statistically complete sample were
  derived from the modeling of the central emission-line widths
  ([\ion{N}{2}] or [\ion{S}{2}]), observed over a $0\farcs25\times
  0\farcs2$ ($R\ltorder 9$ pc) aperture. The experiment has a mean
  detection sensitivity of $\sim 3.9\times10^6\,{\rm M}_\odot$. For
  three sample members with direct determinations of $M_{\rm BH}$ our
  upper limits agree within the uncertainties, while in general our
  upper limits are found to be close to the masses measured in other
  bulges with global properties similar to ours. Remarkably, our
  limits lie quite closely to the recently derived $M_{\rm
  BH}-\sigma_{\star}$ relation. These results support a picture
  wherein the black-hole mass and overall galaxy structure are closely
  linked, as galaxies with exceptionally high $M_{\rm BH}$ at a given
  $\sigma_{\star}$ apparently are rare.
\end{abstract}


%
%
%
\section{Introduction}
\label{sec:UppLim_intro}

The last few years have seen great progress in studying the dark mass
concentrations at the centers of ``ordinary'' quiescent galaxies,
showing that they are very common and demonstrating in some cases that
they must be supermassive black holes (SMBHs) by ruling out all
astrophysically viable alternatives. Indeed, a picture is emerging in
which SMBHs in the range of $10^6$ to $10^9\,\Msun$ are an integral
part of galaxy formation (\eg, Kauffmann \& Haehnelt 2000).
Two examples in the local universe stand out with particularly
convincing evidence as SMBHs: the Milky Way and NGC~4258. At the
Galactic Center, direct observations of individual stars (Genzel \etal
1997; Eckart \& Genzel 1997; Ghez \etal 1998) and a stream of ionized
gas (Herbst \etal 1993) orbiting Sgr~A$^*$ show that all the
dynamically relevant mass inside $\sim 1$ pc, $2.6\times 10^6\,\Msun$,
is concentrated with a density of $\rho >10^{12}\,\Msun\ {\rm
pc}^{-3}$. In NGC~4258, a disk of masing molecular gas is orbiting the
center with a Keplerian rotation curve as traced by H$_2$O maser
emission (Miyoshi \etal 1995); models imply that $M_{\rm BH}\approx
3.6\times 10^7\,\Msun$ and $\rho >4\times10^{9}\,\Msun\ {\rm
pc}^{-3}$.
Alternatives to SMBHs, such as clusters of brown dwarfs or stellar
remnants, can be ruled out in these two cases (e.g. Maoz 1995, 1998).

A number of techniques have matured that have demonstrated the
presence of a central dark mass concentration in an ever growing
number of nearby galaxy nuclei, with mass estimates accurate to a
factor of $\sim$2 and concentration limits of $\rho >10^{6-8}\,\Msun\
{\rm pc}^{-3}$. Simple analogy with the exemplary cases where the SMBH
presence is all but proven, and the connection to active galactic
nuclei (AGN) activity at various intensity levels, suggest strongly
that these central dark masses are SMBHs as well. The most widely used
technique is stellar dynamical modeling (e.g., Dressler \& Richstone
1988; Kormendy et al. 1996, 1997; van der Marel \etal 1997; Cretton \&
van den Bosch 1999; Gebhardt \etal 2000a), which has provided mass
estimates for over two dozen nuclei, mostly in massive, early-type
galaxies. Modeling the kinematics of ionized gas has produced a number
of additional mass measurements (e.g., Harms et al. 1994; Ferrarese,
Ford, \& Jaffe 1996; Bower \etal 1998; Verdoes Kleijn et al. 2000;
Sarzi \etal 2001, hereafter S01; Barth \etal 2001a).  Finally,
application of results from reverberation mapping of active galactic
nuclei have yielded central virial mass estimates for Seyfert galaxies
(Ho 1999; Wandel, Peterson, \& Malkan 1999) and QSOs (Kaspi \etal
2000) that seem to be robust (Gebhardt
\etal 2000b; Ferrarese \etal 2001).

Taken together, these results show that $M_{\rm BH}$ is correlated
both with the stellar luminosity $L_{\rm bulge}$ (Kormendy \&
Richstone 1995; Magorrian \etal 1998; Ho 1999; Kormendy et al. 2001)
and, more tightly, with the velocity dispersion of the bulge
$\sigma_{\star}$ (Gebhardt \etal 2000a; Ferrarese \& Merritt 2000).
By contrast, $M_{\rm BH}$ is unrelated to the properties of galaxy
disks (Kormendy et al. 2001).  The growth of SMBHs appears to be
closely linked with the formation of bulges.  However, the actual
slope and scatter of the $M_{\rm BH}-L_{\rm bulge}$ and $M_{\rm
BH}-\sigma_{\star}$ relations are still under debate. It is also
important to remember that our knowledge about SMBHs is very uneven
across the Hubble sequence of galaxies.  The existing samples are
preferentially weighted toward early-type galaxies with very massive
black holes. From an observational point of view, there is a pressing
need to acquire better $M_{\rm BH}$ statistics for spiral galaxies.

Motivated by the recent progress and the emerging correlations, but
also by the desire to improve the black hole census in spirals, we
derive mass constraints on SMBHs potentially present in the bulges of
sixteen nearby disk galaxies. As we will show, these constraints
constitute significant progress, both in terms of the number of target
galaxies and in terms of broadening the range of parent galaxies with
significant constraints.

We draw on spectra of nearby nuclei obtained with the Space Telescope
Imaging Spectrograph (STIS) onboard the {\it Hubble Space Telescope
(HST)}, taken as part of the Spectroscopic Survey of Nearby Galaxy
Nuclei (SUNNS) project (Shields et al. 2000; Ho \etal 2000; S01; Rix
et al. 2001). Only four of our original twenty-four target galaxies
showed extended line emission with symmetric kinematics and hence were
suited for a direct $M_{\rm BH}$ determination. These cases have been
modeled by S01 and yielded $M_{\rm BH}$ estimates. Here we analyze
data for sixteen galaxies from SUNNS with central [\ion{N}{2}]
$\lambda\lambda$6548, 6583 and [\ion{S}{2}] $\lambda\lambda$6716, 6731
line emission from ionized gas. In most cases the central gas emission
is spatially unresolved, or only marginally so, and has line widths
$\sim 10^{2}$~\kms.

Spatially unresolved lines do not permit precise mass estimates, but
potentially do allow us to derive useful upper limits on $M_{\rm BH}$.
This is because the line emission in our central aperture must arise
from gas at a distance from the center that is, at most, equal to the
physical dimension of the region subtended by the central aperture
itself, for our sample typically $\sim 9$ pc.
If the gas motions are orbital, all velocities and hence the
integrated line width will scale as $\sqrt{M_{\rm BH}}$. Note that the
resulting limit on $M_{\rm BH}$ scales linearly with the central
aperture size, affording {\it HST}\ an order of magnitude gain over
ground-based observations (e.g., Salucci \etal 2000), and making the
derived limits astrophysically interesting.

At parsec-scale distances from galactic centers the gas is subject not
only to gravitational forces but also to gas pressure and magnetic
forces (e.g., for the Milky Way, Timmermann \etal 1996; Yusef-Zadeh,
Roberts, \& Wardle 1997). In general, these other effects cause
additional line broadening.

The paper is organized as follows.  In \S
\ref{sec:UppLim_Obser&DataRed} we present the spectroscopic and
photometric STIS observations, and in \S
\ref{sec:UppLim_LinewidthsModelling} we describe our modeling of the
ionized gas kinematics. In \S \ref{sec:UppLim_results} and \S
\ref{sec:UppLim_Disc&Concl}, we present our results and draw our
conclusions.

%
%
%
\section{Observations and Data Reduction}
\label{sec:UppLim_Obser&DataRed}

\subsection{Galaxy Sample}
\label{subsec:UppLim_GalSample}

All sample galaxies, mostly early-type disk galaxies (S0 -- Sb), were
observed with STIS as part of the SUNNS project; the full details of
this program will be reported elsewhere (Rix \etal 2001). The SUNNS
galaxies are drawn from the Palomar spectroscopic survey of nearby
galaxies (Filippenko \& Sargent 1985; Ho, Filippenko, 
\& Sargent 1995, 1997) and include all S0 to
Sb galaxies within 17~Mpc known to have line emission ($\gtrsim
10^{15}$ ergs s$^{-1}$ cm$^{-2}$) within a $2\asec \times 4\asec$
aperture. The present sample constitutes the subset of SUNNS objects
with sufficient central line flux in H$\alpha$ or [\ion{N}{2}] to
provide adequate signal-to-noise (S/N $\gtrsim 10$). With the
experimental setup described below, the sample is effectively defined
by galaxies that have H$\alpha$ or [\ion{N}{2}] line fluxes $\gtrsim
10^{-14}$ ergs s$^{-1}$ cm$^{-2}$ within a $0\farcs25 \times 0\farcs2$
central aperture, which correspond to line luminosities of $\gtrsim
3\times 10^{38}$ ergs s$^{-1}$ at the mean sample distance of 15~Mpc.
%
Our sample is statistically well defined and complete in the sense
that the original SUNNS sample was a volume-limited sample selected by
emission-line flux within a (much larger) $2\asec \times 4\asec$
aperture.
%
The basic parameters of the target galaxies are given in Table
\ref{tab:UppLim_GalSample}.

\subsection{Observations}
\label{subsec:UppLim_Observations}

{\sl HST}\ observations were acquired for all objects in SUNNS during
1998 and 1999.  We placed the $0\farcs2 \times 52\asec$ slit across
each nucleus along an operationally determined position angle, which
is effectively random with respect to the galaxy orientation.
After initial 20-s acquisition exposures with the optical long-pass
filter (roughly equivalent to $R$), from which we derive surface
photometry of the central regions, three exposures totaling
approximately 45 minutes were obtained with the G750M grating; this
resulted in spectra that cover 6300 \AA\ to 6850 \AA\ with a full-width
at half maximum resolution for extended sources of 1.6 \AA.

For 9 of the 16 galaxies in the present sample, the telescope was
offset by $0\farcs05$ ($\sim$1 pixel) along the slit between repeated
exposures to aid in the removal of hot pixels and cosmic rays.  The
two-dimensional (2-D) spectra were bias- and dark-subtracted,
flat-fielded, aligned, and combined into single frames. Cosmic rays
and hot pixels remaining in the combined 2-D spectra were cleaned
following the recipe of Rudnick, Rix, \& Kennicutt (2000). The 2-D
spectra were then corrected for geometrical distortion and then
wavelength and flux calibrated with standard {\scriptsize STSDAS}
procedures within {\scriptsize IRAF}\footnote{IRAF is distributed by
the National Optical Astronomical Observatories, which are operated by
AURA, Inc. under contract to the NSF.}

To represent the generic ``nuclear spectrum'' of each galaxy, we
extracted aperture spectra five pixels wide ($\sim 0\farcs25$),
centered on the brightest part of the continuum. The acquisition
images indicate that the uncertainties in the galaxy center due to
dust are $\ltorder 0.25$ pix $\sim 0\farcs012$ (see also S01). In
essence, therefore, the extracted spectra represent the average
central emission, convolved with the STIS spatial point-spread
function (PSF) and sampled over an aperture of $0\farcs25 \times
0\farcs2$, or 18~pc $\times$ 14~pc for the mean sample distance of
15~Mpc.

For three of our galaxies, central stellar velocity dispersions were
either not available in the literature (NGC~3992 and NGC~4800) or
quite uncertain (NGC 3982; Nelson \& Whittle 1995).
Therefore, we obtained new spectroscopic data for these objects on two
observing runs: with the Boller \& Chivens spectrograph at the Bok
90-inch telescope in May 2000 for NGC~3992, and with the Double
Spectrograph (Oke \& Gunn 1982) at the Palomar 200-inch telescope in
June 2001 for NGC~3982 and NGC~4800.
At Kitt Peak we used the 600 grooves mm$^{-1}$ grating to cover
3600--5700 \AA\ with a pixel scale on the CCD of 1.86 \AA\ pix$^{-1}$,
while for the Palomar run we observed the Ca infrared triplet using
the 1200 grooves mm$^{-1}$ grating on the red side of the Double
Spectrograph, with a pixel scale of 0.63 \AA\ pix$^{-1}$.
Spectra were extracted for apertures of $3\farcs3 \times 2\farcs5$ and
$3\farcs7 \times 2\farcs0$ for the Bok and Palomar spectra,
respectively.
Total exposure times were 40, 30, and 60 minutes for NGC~3982,
NGC~3992, and NGC~4800, respectively. The stellar velocity dispersions
were measured following the method of Rix \etal (1995), and the
obtained values are reported in Table \ref{tab:UppLim_GalSample}.

\subsection{Central Emission-Line Widths and Flux Profiles}
\label{subsec:UppLim_CentralLinewidthsAndFluxProfiles}

To quantify the emission-line velocity widths in these nuclear
spectra, we simultaneously fit Gaussians of single width
$\sigma_{cen}$ to the [\ion{N}{2}] $\lambda\lambda$6548, 6583 and
[\ion{S}{2}] $\lambda\lambda$6716, 6731 emission-line doublets, using
the {\scriptsize IRAF} task {\scriptsize SPECFIT}. The profiles of
the [\ion{N}{2}] and [\ion{S}{2}] lines were found to be roughly
Gaussian.
We restricted ourselves to line widths from forbidden transitions to
side-step the impact of a possible broad (broad-line region) component
arising from radii much smaller than our observational aperture.
However, in all objects where only a narrow H$\alpha$ emission-line
component was present, we also included H$\alpha$ in the fit.
In the few cases with prominent, very broad H$\alpha$ lines (\eg,
NGC~4203, Shields \etal 2000; NGC~4450, Ho \etal 2000), particular
care was taken to minimize the impact of the very broad lines on the
estimate of $\sigma_{cen}$ in the adjacent [\ion{N}{2}] lines.
In virtually all objects the S/N is in excess of 10, and hence the
formal errors in the estimated line width are negligible for the
subsequent analysis.
The instrumental line width derived from comparison lamps is $\sigma_{inst}
\approx 32$ \kms\ and was subtracted from the raw measurement of
$\sigma_{cen}$ in quadrature; for all but two objects these line width
corrections were negligible, implying that the intrinsic widths were
well resolved.
This correction also spares us from accounting in the following
modeling of the line width for the broadening due to the instrumental
line spread function.
The resulting values for $\sigma_{cen}$ are listed in Table 1.  Their
characteristic errors, including the correction for the instrumental
line width, are less than 10 \kms.

As we will detail below, any information on the gas spatial flux
distribution on scales $\lesssim 0\farcs25$ provides a valuable
constraint for the central line width modeling procedure.  Therefore,
we also obtained radial profiles along the slit direction for the
ionized gas flux, by fitting a Gaussian to the [\ion{N}{2}]
$\lambda$6583 line-flux profile along the spatial direction on the 2-D
spectra.
We chose [\ion{N}{2}] because among our sample galaxies this is almost
always the brightest line, and because it is less likely to be
affected by underlying absorption features in the stellar continuum
than H$\alpha$. The [\ion{N}{2}] emission-line flux profiles are shown
in Figure \ref{fig:UppLim_FluxProfilesAndFitAndContinuum}.

\figureone

%
%
%
\section{Modeling the Central Line Width}
\label{sec:UppLim_LinewidthsModelling}

\subsection{Basic Concept}
\label{subsec:UppLim_BasicIdea}

We are now faced with converting the observed central line widths into
estimates for the central black-hole mass.

To start, we assume that the ionized gas motion is dominated solely by
gravity. In this case the central line width depends on: (a) the total
gravitational potential of the putative SMBH and of the surrounding
stars; (b) the spatial emissivity distribution (\eg, that of a disk
inclined at $\cos{i}$); and (c) the ``kinematic behaviour'' of the
ionized gas, for example ``dynamically cold'' gas moving on circular
orbits or hotter gas with hydrostatic support.
The lack of spatially resolved information on the gas flux
distribution within the central $0\farcs25 \times 0\farcs2$ aperture
means that we can only derive upper limits to $M_{\rm BH}$; if the
emission-line flux within the aperture arose from $R \ll R_{aperture}
\approx 0\farcs1$, arbitrarily small values of $M_{\rm BH}$ could
explain the observed line width.
If the gas motion is also affected by non-gravitational forces, such
as outflows, magnetic fields, or supernova winds, this would broaden
the integrated line velocity width additionally, and hence lower the
required black-hole mass needed to explain a given $\sigma_{cen}$.
By ignoring non-gravitational forces we are therefore conservative in
estimating upper limits for the central black-hole mass.
The absence of constraints on the importance of non-gravitational
forces constitutes a second reason (besides the lack of spatially
resolved information on the gas flux) why it is not possible to secure
the presence of a SMBH in our sample galaxies, as hypothetically the
observed line widths could be entirely explained by non-gravitational
effects.

If the functional form of the potential well is fixed, then the
central line width will scale with the potential for any given choice
of the emissivity distribution and for the gas kinematical behaviour.
In the simplest case of a purely Keplerian potential induced by a
SMBH, $\Phi_{\rm BH}$, the expected central line width will scale as
the square root of the black-hole mass.
As the circular velocity at any given reference radius $R_{ref}$,
$v_c(R_{ref})$, will scale in the same way, the ratio between
$\sigma_{cen}$ and $v_c(R_{ref})$ is independent of black-hole mass.
The task at hand is therefore to derive a plausible range of values
for this ratio by varying the spatial emissivity distribution, and
then to obtain a mass range for the putative SMBH from the observed
central line width via $\sigma_{cen} \rightarrow v_c^2(R_{ref})
\rightarrow M_{\rm BH}= v_c^2(R_{ref})R_{ref}/G$.

The same would hold for a sequence of purely stellar potentials
derived from the luminosity density with differing mass-to-light
ratios $\Upsilon$.
When both the stellar and the SMBH contribution to the gravitational
potential are considered, the shape of the rotation curve, and hence
$\sigma_{cen}/v_c(R_{ref})$, will depend on the relative weight of
$M_{\rm BH}$ and $\Upsilon$.

In all cases we will proceed through the following steps in order to
make a prediction for the gas velocity dispersion within the central
aperture:

\begin{itemize}
\item Specify the spatial gas emissivity distribution and the
  gravitational potential and choose the kinematic behaviour of the
  gas.
\item Construct 2-D maps for the moments of the line-of-sight velocity
  distribution (LOSVD) at any position $(x,y)$ on the sky
  \begin{equation} \overline{\Sigma v^k}(x,y) =\int\,{\rm
    LOSVD} \,(x,y,v_z)v_z^kdv_z \,\,(k=0,1,2), 
  \label{eq:LOSVDmoments}
  \end{equation}
  as they would appear without the limitations of the spatial
  resolution; the first moment, for instance, is the mean gas
  velocity.
\item Convolve each of the 2-D $\overline{\Sigma v^k}$
  maps with the STIS PSF.
\item Sample the convolved $\overline{\Sigma v^k}_{conv}$ 2-D maps over
  the desired aperture to obtain the PSF-convolved, aperture-averaged
  LOSVD velocity moments $\overline{\Sigma v^k}_{conv,ap}$, which are
  directly comparable to the observables.
\item In particular, compute the ionized gas flux $f_{ap}$, the
  projected mean streaming velocity $\overline{v}_{ap}$, and the
  velocity dispersion $\sigma_{ap}$ within the desired aperture
  through $f_{ap} = \overline{\Sigma v^0}_{conv,ap}$,
  $\overline{v}_{ap} = \overline{\Sigma v^1}_{conv,ap}\,/f_{ap}$, and
  $\sigma_{ap} = \sqrt{\,\overline{\Sigma v^2}_{conv,ap}\,/f_{ap} -
  \overline{v}_{ap}^2}$, respectively. The last quantity,
  $\sigma_{ap}$, can be compared with the measured velocity width
  $\sigma_{cen}$.
\end{itemize}

In what follows, we will first derive upper limits on $M_{\rm BH}$
assuming that the gas is moving on circular orbits in a coplanar,
randomly oriented disk within a Keplerian potential (\S
\ref{subsec:UppLim_TheDiskModelling}); then we consider the impact of
the stellar potential on this disk modeling (\S
\ref{subsec:UppLim_TheStarContribution}); finally, we will explore a
seemingly very different situation for the kinematical behaviour of the
gas, that of hydrostatic equilibrium (\S \ref{subsec:UppLim_HydroEq}),
to demonstrate that our results are robust with respect to the
underlying model assumptions. We anticipate that the most conservative
upper limits on $M_{\rm BH}$ are derived from the first approach,
where the impact of the stellar potential is neglected.

\subsection{The Keplerian Disk Modeling}
\label{subsec:UppLim_TheDiskModelling}

We start with the simple and plausible assumption that the ionized gas
moves on circular orbits at the local circular velocity, which in turn
is dictated solely by the gravitational influence of the putative
SMBH, $v^2_c(R) = GM_{\rm BH}/R$. We further assume that the gas
resides in a coplanar disk of unknown inclination with an
intrinsically axisymmetric emissivity distribution, $\Sigma(R)$,
centered on the stellar nucleus.
Our best guess for $\Sigma(R)$ is derived from the data themselves.
In this Keplerian disk the LOSVD at each position $(x,y)$ on the sky
plane is just
\begin{equation} 
 {\rm LOSVD} \,(x,y,v_z)=\Sigma_{proj}(x,y)\, \delta[v_{c,proj}(x,y)-v_z]
\label{eq:LOSVDkeplerian}
\end{equation}
and its $\overline{\Sigma v^k}$ velocity moments are simply given by
$\Sigma_{proj}(x,y)$, $\Sigma_{proj}(x,y)\,v_{c,proj}(x,y)$, and
$\Sigma(x,y)\,v^2_{c,proj}(x,y)$, respectively, where $\Sigma_{proj}$
and $v_{c,proj}$ are the projected gas surface brightness and circular
velocity.
To deal with the central circular velocity singularity, we neglected
the contribution of the central point of the 2-D maps for the
$\overline{\Sigma v^k}$ velocity moments, and we refined their grid
sizes until no further substantial increase in the predicted line
widths was found. The adopted grid size in our models corresponds to
$0\farcs005$, or one tenth of a pixel.

The geometry of the projected velocity field will depend on the disk
orientation, specified by its inclination $i$ with respect to the sky
plane and its major axis position angle $\phi$ with respect to the
slit direction.
We have no information on the gas disk orientation within our central
aperture, since the dust-lane morphology we employed in S01 cannot be
used to provide a constraint on such small scales.
Therefore, we need to explore all possible disk orientations to derive
the probability distribution for $\sigma_{ap}/v_{c}(R_{ref})$, the
ratio between the predicted central velocity dispersion and the
circular velocity at the reference radius.
We adopt $R_{ref}=0\farcs125$, corresponding to the distance of the
central aperture edge from the center along the slit direction.
We cover the possible disk orientations by constructing a grid of
models with equally spaced $\cos{i}$ and $\phi$.

For the intrinsic radial surface brightness profile of the gas we
assumed a Gaussian
\begin{equation}
 \Sigma(R)=a_{flux}\,e^{-R^2/2\sigma^2_{flux}},
\label{eq:intrinsicflux}
\end{equation}
and we derived $a_{flux}$ and $\sigma_{flux}$ by matching the observed
emission flux profiles along the slit (see \S
\ref{subsec:UppLim_CentralLinewidthsAndFluxProfiles}). This match
again involves convolving the intrinsic $\Sigma(R)$ with the STIS PSF,
which we parameterized as a sum of Gaussian components (see S01).
The choice of a Gaussian for the intrinsic surface brightness
distribution was a matter of convenience, for the convolution process
in this case is simply analytical. Intrinsically more concentrated
profiles, such as exponential ones, would also reproduce the data once
convolved with the STIS PSF, and would lead to tighter upper limits on
$M_{\rm BH}$, which make our choice more conservative.
We fit only the central five flux pixels for each galaxy, corresponding
to the same region subtended by our central aperture.

Figure \ref{fig:UppLim_FluxProfilesAndFitAndContinuum} displays our
best fits of $\Sigma(R)$, and illustrates that our model
(Eq. \ref{eq:intrinsicflux}) matches the data well within
$<0\farcs125$ in all 

\figuretwo

\noindent
cases.\footnote{
  All [\ion{N}{2}] flux profiles of Figure
  \ref{fig:UppLim_FluxProfilesAndFitAndContinuum} were symmetrical
  even outside the central aperture region, with the noticeable
  exception of NGC~4501 and NGC~4698.
  This last Sa galaxy shows the presence of a stellar (Bertola \etal
  1999) and gaseous (Bertola \& Corsini 1999) core with an angular
  momentum perpendicular to that of the main galactic disk. This
  core can be identified as a disk from {\it HST}\ imaging (Scarlata \etal
  2001).
  Consistent with these findings, a recent accretion event may
  explain why NGC~4698 is one of the two galaxies, among our sample of
  16, to exhibit a strongly asymmetric gas distribution within
  $0\farcs3$ ($\pm 6$ rows) where, assuming that the nuclear regions
  of our galaxies are dominated by SMBHs with masses
   consistent with the $M_{\rm BH}-\sigma_{\star}$ relation, the dynamical
  timescale ranges from 0.5 to 12.0 Myr.}
Table \ref{tab:UppLim_Results} lists the
best-fitting $\sigma_{flux}$ values.
The actual position of the data points with respect to the center of
the fitting function can be explained entirely by a small displacement
($\ll R_{ref}$) of the slit center from that of the gaseous disk,
without violating our assumption of an axisymmetric emissivity
distribution.
Including this small offset in the modeling produces only
negligible variations in the predicted central $\sigma_{ap}$.
In this case the predicted mean velocities $\overline{v}_{ap}$,
although non-zero, are very small and consequently the velocity
dispersions $\sigma_{ap}$ are almost equal to the ones obtained with
perfectly centered apertures (when $\overline{v}_{ap} \equiv
0$).

For comparison, we also show the stellar surface brightness profiles
in Figure \ref{fig:UppLim_FluxProfilesAndFitAndContinuum}, derived
from the stellar continuum in the spectra; this comparison justifies
our assumption that the gas and the stellar distribution are
concentric.
Indeed, an independent fit to the stellar profiles with the same
functional profile adopted for the gas ones led to a mean offset
along the slit direction of only $0.08\pm0.17$ pixels, consistent with
no offset at all.


As an intermediate result of our modeling, we show in Figure
\ref{fig:UppLim_SigOverVcExample} the $\sigma_{\rm ap}/v_{c}(R_{ref})$
ratios obtained in a Keplerian potential for different disk
orientations and a range of typical values for $\sigma_{flux}$; the
predicted $\sigma_{ap}$ does not depend on the total flux, or
$a_{flux}$.
The predicted central line width obviously increases from face-on to
edge-on systems. At a given disk orientation, models with
intrinsically more concentrated gas emissivity always have a larger
line width than those with more extended flux distributions because the gas
resides at smaller radii.
Since the central $0\farcs25 \times 0\farcs2$ aperture is nearly
square, the impact of the position angle parameter $\phi$ on the final
confidence limits for the $\sigma_{ap}/v_{c}(R_{\rm ref})$ ratio is
negligibly small.

The flux distributions in our sample galaxies are concentrated
enough that the predicted line widths are {\it monotonically}

\figurethree

\noindent
decreasing with increasing $\cos{i}$.
Since randomly oriented disks have uniformly distributed $\cos{i}$, we
can use Figure \ref{fig:UppLim_SigOverVcExample} to derive the median
values and $68\%$ upper and lower confidence limits for the $M_{\rm BH}$,
respectively, by simply taking the values of $\sigma_{ap}/v_{c}(R_{\rm
  ref})$ for the models with $\cos{i}$ = 0.5, 0.84, and 0.16.

There is a minor practical complication in this modeling: while
changing the orientation of a disk with a fixed intrinsic flux
distribution, the predicted flux profile along each of the five
$0\farcs05 \times 0\farcs2$ apertures within the central $0\farcs25
\times 0\farcs2$ aperture changes, eventually becoming inconsistent
with the observed one.
Hence, at any given disk orientation we must readjust the
intrinsic flux concentration $\sigma_{flux}$ in order to match the
central five flux data points.
Such a correction is particularly important for highly inclined disks,
which have very different flux profiles when considered at different
position angles.  Since our $M_{\rm BH}$ upper limits are derived for
nearly face-on orientations, these corrections will not strongly
affect our results.
Furthermore, tests have shown that the induced scatter in the
$\sigma_{ap}/v_{c}(R_{ref})$ ratio at any given inclination is
considerably smaller than the face-on to edge-on variation.


For simplicity, we only used a statistical correction for this effect.
For a set of intrinsic values $\sigma_{flux,in}$, we collected all the
central flux profiles predicted for a uniform grid in $\cos{i}$ and
$\phi$. Then we treated each of these profiles as observed ones and
matched each of them with a PSF-convolved Gaussian profile to get a
distribution of $\sigma_{flux,out}$ values and a median
$\langle\sigma_{flux,out}\rangle$ value for each $\sigma_{flux,in}$.
By comparing the median $\langle\sigma_{flux,out}\rangle$ values with
the corresponding $\sigma_{flux,in}$ values
(Fig. \ref{fig:UppLim_SigFluxGuessCorrection}), we can correct for
each galaxy our initial guess of the intrinsic $\sigma_{flux}$,
derived by matching the observed central flux profiles of 

\figurefour

\noindent
Figure
\ref{fig:UppLim_FluxProfilesAndFitAndContinuum}.
For a given galaxy the corrected flux concentration to be input into
the models is characterized by the $\sigma_{flux,in}$ value that,
according to the previous scheme, lead to a median
$\langle\sigma_{flux,out}\rangle$ equal to the $\sigma_{flux}$ of that
galaxy.
This $\sigma_{flux,in}$ value describes the intrinsic flux
distribution that, when considering different disk orientations, leads
to the predicted flux profiles that are the most consistent with the
observed one for the given galaxy.

We call these particular $\sigma_{flux,in}$ for each galaxy the
corrected $\sigma_{flux,corr}$ values, which we list in Table
\ref{tab:UppLim_Results} along with the ($+1\sigma$) upper limits on
$M_{\rm BH}$ obtained adopting them.

As a final remark we notice that at the mean distance of our sample
galaxies ($\sim 15$ Mpc) and considering the typical mass of their
central SMBH's ($\sim 2.4\times 10^7 {\rm M}_\odot$, as predicted by
the $M_{\rm BH}-\sigma_{\star}$ relation), the adopted intrinsic flux
distributions are generally concentrated enough that a double-horned
LOSVD should be expected, expecially for highly inclined disks.
The fact that such a feature is not found in the observed
emission-line profiles could represent evidence of an intrinsic
turbulence in the gas.
Indeed, when the full LOSVD is properly constructed (by collecting at
each velocity bin the total flux within the $0\farcs25 \times
0\farcs2$ aperture that arises from the corresponding iso-velocity
slice of the flux distribution, once convolved with the STIS PSF), the
double-horned shape disappears when an intrinsic gas velocity
dispersion is introduced.
In the favorable case, within this context, of a nearly face-on disk
with $\cos{i}=0.84$, a velocity dispersion of about $\sim 30$ \kms\
would be required on average to smooth the double-horned shape, thus
increasing the predicted line widths by 9\% and decreasing the derived
upper-limits on $M_{\rm BH}$ by 19\%.
Too many assumptions will be required to make this correction on the
case by case basis, and would probably require direct fitting of the
observed emission lines, as done by Barth \etal (2001b). By adopting
our current approach, our limits on $M_{\rm BH}$ remain conservative
upper-bounds.

\subsection{The Stellar Contribution}
\label{subsec:UppLim_TheStarContribution}

We now proceed to evaluate the impact of the stellar potential
$\Phi_{\star}$ on our modeling.
We start by estimating for each galaxy the expected radius of the
``sphere of influence'' of the SMBH, $r_{infl}=GM_{\rm
BH}/\sigma_{\star}^2$, within which $M_{\rm BH}$ dominates the
dynamics of a galaxy with a stellar velocity dispersion
$\sigma_{\star}$.
For this estimate we adopt the $M_{\rm BH}-\sigma_{\star}$ relation as
parameterized by Gebhardt \etal (2000), $\log M_{\rm BH} =
3.75\log\sigma_{\star} -0.55$. Then, $\log r_{infl} =
1.75\log\sigma_{\star} -2.92$, where $M_{\rm BH}$ is in units of ${\rm
M}_{\odot}$, $\sigma_{\star}$ in km s$^{-1}$, and $r_{infl}$ in pc.
In Figure \ref{fig:UppLim_Rinfluence} we compare $r_{infl}$ with the
physical scale corresponding to the mean radius of our central
aperture $R_{aperture}=\sqrt{(0\farcs25 \times 0\farcs2)/\pi}\approx
0\farcs13$, and find that for many of our galaxies $r_{infl}
\leq R_{aperture}$, indicating that the stellar mass $M_{\star}$
within $R_{aperture}$ is comparable to or exceeds $M_{\rm BH}$.
Therefore, our analysis needs to account for the stellar mass, and we
need to derive the stellar mass density profiles $\nu_{\star}(r)$, in
particular for galaxies with smaller $\sigma_{\star}$.

The relative importance of $M_{\star}$ and $M_{\rm BH}$ on the central
line width depends also on the spatial extent of the gas emissivity.
In particular even when $r_{infl} < R_{aperture}$, the presence of the
SMBH can still be noticed from the observed $\sigma_{cen}$ if
$\sigma_{flux} \approx r_{infl}$ --- that is, if most of the collected
flux within $R_{aperture}$ was emitted by gas moving on nearly
Keplerian orbits.
Further, the impact of $\Phi_{\star}$ on the predicted $\sigma_{ap}$
also decreases with increasing flux concentration since in general the
circular velocity curve due to the stars increases monotonically with
the distance from the galactic center. Indeed the central slope of
stellar densities profiles can be represented by a power-law
$\nu_{\star}(r)\sim r^{-\alpha}$ with $\alpha \leq 2$ (Gebhardt \etal
1996).

Figure \ref{fig:UppLim_Rinfluence} shows $\sigma_{flux}$ is in general
smaller than $R_{aperture}$, and we expect that the inclusion of the
stellar mass in our modeling will cause only a modest correction in
the black-hole masses inferred in \S
\ref{subsec:UppLim_TheDiskModelling}.

To quantify the stellar mass contribution in each galaxy, we derived
the mass density profile $\nu_{\star}(r)$ by deprojecting the stellar
surface brightness distribution $\Sigma_{\star}(R)$ obtained from the
STIS acquisition image, assuming spherical symmetry and a constant
mass-to-light ratio $\Upsilon$.
We applied the same multi-Gaussian algorithm adopted in S01 to
circularly averaged $\Sigma_{\star}(\sqrt{ab})$ surface brightness
profiles, extracted using the {\scriptsize IRAF} task {\scriptsize
ELLIPSE} and color corrected into Johnson $R$-band magnitudes using
the {\scriptsize IRAF} package {\scriptsize SYNPHOT} and assuming
E--S0 galaxy templates.
Gaussian components with $\sigma \leq 0.5$ pixel were considered
as unresolved point sources and hence were excluded from the
stellar mass budget.
For simplicity we adopted for all galaxies $\Upsilon=5\; {\rm
M}_{\odot}/{\rm L}_{\odot}$, rescaled from van der Marel (1991) for
$H_0=75$ km s$^{-1}$ Mpc$^{-1}$, instead of deriving individual values for
$\Upsilon$ by matching ground-based $\sigma_{\star}$ measurements (see
S01).


Including the stellar potential in the modeling results in a 
27\% reduction of the median black-hole masses (for $\cos i = 0.5$)
needed to explain the observed central line widths.
Nonetheless, this effect is important for galaxies with small
$\sigma_{\star}$ values; the median $M_{\rm BH}$ decreased by 37\% for
the sample galaxies with the lowest $\sigma_{\star}$ (NGC~3982,
NGC~4321, and NGC~4548), but only by 3\% for the ones with the highest
$\sigma_{\star}$ (NGC~4143).

The impact of the stellar mass on the upper limits of the black-hole
mass, which are central to our analysis, is yet smaller, on average
less then 12\% (see Tab. \ref{tab:UppLim_Results}).
Indeed, similar to the purely Keplerian case, the predicted central
line widths are always found to be increasing monotonically with 
disk inclination, such that the derived $+1\sigma$ upper limits on
$M_{\rm BH}$ are obtained here also from models with nearly face-on disks.
In this situation the circular velocities needed by the model to
explain the observed line widths always by far exceed the ones
provided only by the stellar potential.

\figurefive

\subsection{Gas in Hydrostatic Equilibrium}
\label{subsec:UppLim_HydroEq}

So far we have assumed that the observed central line widths are due
to pure orbital motion of gas in a disk around the galaxy center.
But even if we ignore non-gravitational effects, we still need to
investigate how much the derived $M_{\rm BH}$ upper limits depend on
different choices for the kinematic behaviour of the ionized gas.
In particular, it is conceivable that the gas is at least in part
supported by gas pressure. In order to explore the impact of such
pressure on our results we considered the extreme case of pure
hydrostatic support.
For any emissivity density profile $\rho$, we need only the second
{\scriptsize LOSVD} velocity moment $\overline{\Sigma v^2}$ without
streaming motions. We obtain this moment by solving the hydrostatic
equilibrium equation $d(\rho\sigma^2)/\rho dr=-d\Phi/dr$ for the gas
velocity dispersion $\sigma(r)$, and then by integrating the
luminosity-weighted $\sigma$ along the line of sight.

Under these assumptions, we proceeded to match the observed
$\sigma_{cen}$ for all the galaxies, assuming Gaussian profiles for
their emissivity densities $\rho$, whose width has to match the
observed central flux profiles. We thus obtained for each galaxy a
value for the mass of the putative SMBH, in a purely Keplerian
potential.

From this exercise we found that the predicted central velocity
dispersions $\sigma_{ap}$ correspond to values from the rotating disk
model with disk inclinations always between 0.65 and 0.71.
Consequently, the $M_{\rm BH}$ values inferred by considering hydrostatic
equilibrium lie within the $\pm 1\sigma$ confidence limits 
obtained from the rotating disk models, and all previous results
hold.

%
%
%
\section{Results}
\label{sec:UppLim_results}

We have explored the dynamical implications of the observed
emission-line widths arising from the central $\sim 10$ pc of our
sample galaxies, and we have demonstrated that the most conservative
$1\sigma$ upper limits of $M_{\rm BH}$ are obtained assuming that the
gas resides in a nearly face-on disk ($i\sim 33^{\circ}$;
$\cos{i}=0.84$) moving in circular orbits around a central
SMBH. Fortunately, extremely face-on orientations are statistically
rare.

In Figure \ref{fig:UppLim_UppLimKeplerian} we place our $M_{\rm BH}$ upper
limits derived for the Keplerian case (\S
\ref{subsec:UppLim_TheDiskModelling}) in the $M_{\rm BH}-\sigma_{\star}$
plane .
For a comparison with the $M_{\rm BH}-\sigma_{\star}$ relation (as
parameterized by Gebhardt \etal 2000a) we need to scale each
$\sigma_{\star}$ to $\sigma_{e}$, the value that would have been
observed within a circular aperture of $R = R_e$.
The proper computation of such a quantity, which should include the
contribution from the rotation, was not possible for all the galaxies
in our sample, since the necessary combination of surface brightness,
velocity dispersion, and rotation velocity radial profiles was not
available from the literature for all objects.
The only choice in this case is to use the algorithm of J{\o}rgensen,
Franx, \& Kjaergaard (1995), which was derived on the basis of
kinematical data for 51 elliptical and lenticular galaxies and which
also accounts for the effect of rotation.
For the effective radii we adopted the seeing-corrected values from
Baggett, Baggett \& Anderson (1998). In the few cases where this last
compilation did not provide $R_e$ measurements (NGC~3982 and
NGC~4800), we assumed $\sigma_{e} = \sigma_{\star}$.

Figure \ref{fig:UppLim_UppLimKeplerian} shows that, with one exception
(NGC 4143), our derived upper limits on $M_{\rm BH}$ lie within the
scatter of the $M_{\rm BH}-\sigma_{\star}$ relation, or above it. In
this sense, the proposed $M_{\rm BH}-\sigma_{\star}$ relation passes
this observational test for more than a dozen objects.
Furthermore, except for the three galaxies with the lowest value of
$\sigma_{\star}$, our upper limits on $M_{\rm BH}$ exceed the values
predicted by the $M_{\rm BH}-\sigma_{\star}$ relation only by a modest
amount (in average by a factor of $\sim 4.6$).


In particular for the galaxies in our sample with actual $M_{\rm BH}$
measurements from spatially resolved kinematics (S01), the present upper limits
are consistent, within the errors, with the published $M_{\rm BH}$ values.
Moreover, when our $+1\sigma$ upper limits are compared with other
published $M_{\rm BH}$ measurements (as compiled recently by Kormendy
\& Gebhardt 2001, Fig. \ref{fig:UppLim_UppLimKeplerian_G00_Mbhs}), most appear 
to be quite close to the actual values for the black-hole mass, with
the exception of the same galaxy lying below the $M_{\rm
BH}-\sigma_{\star}$ relation (NGC~4143) and the three objects in our
sample with the lowest value of $\sigma_{\star}$ (NGC~3982, NGC~4321,
and NGC~4548).
Figure \ref{fig:UppLim_UppLimKeplerian_G00_Mbhs} therefore shows that
for the bulk of our sample with $100\,{\rm km\,s^{-1}} \leq
\sigma_{\star} \leq 200\,{\rm km\,s^{-1}}$ our line-width modeling
technique gives results statistically consistent with the values
obtained through other techniques. 
The current data do not provide any indication that the spiral and
lenticular galaxies in our sample differ in their black-hole masses.
Further, our upper limits on $M_{\rm BH}$ do not seem to differ between
barred or unbarred host galaxies, or as a function of the nuclear
spectral classification.


For any interpretation of upper limits, the basic sensitivity of the
experiment is crucial. Our $M_{\rm BH}$ sensitivity limit does not only
depend on the physical size of the resolution element and on the
amount of stellar mass, but also on the spatial emissivity
distribution of the gas.
Indeed, we often find the spatial extent of the ionized gas to be
smaller than the dimension of the $0\farcs25 \times 0\farcs2$
aperture.
Considering the spatial emissivity of each of our sample galaxies, we
can derive conservative sensitivity limits for a nearly face-on disk
($\cos{i}=0.84$) by first computing the predicted line widths arising
in a purely stellar potential, and then by asking what values of
$M_{\rm BH}$ need to be added in order to increase those line widths
by their typical measurement error, conservatively $\sim$10 \kms\ (see
\S \ref{subsec:UppLim_CentralLinewidthsAndFluxProfiles}).

\figuresix

With a mean value of 3.9$\times10^6$ M$_{\odot}$, the derived
sensitivity limits lie well below the inferred $+1\sigma$ upper limits
on $M_{\rm BH}$ (see Fig. \ref{fig:UppLim_UppLimKeplerian} and Table
\ref{tab:UppLim_Results}), which can therefore be considered robust.
We notice that by assigning a constant mass-to-light ratio to
{\it all\/} Gaussian components of the luminosity density profile,
including the ones with $\sigma \leq 0.5$ pixel (see \S
\ref{subsec:UppLim_TheStarContribution}), the derived sensitivity
limits increase only by 30\%, to a mean value of 5.1$\times10^6$
M$_{\odot}$.

Our $+1\sigma$ upper limits for the whole sample correspond to
$M_{\rm BH}$ values produced by nearly face-on disks ($\cos{i}=0.84$).  As
mentioned, we find one object, NGC~4143, for which the derived
upper limit on $M_{\rm BH}$ is below the $M_{\rm BH}-\sigma_{\star}$ relation.
For 16 sample members, the expected number of disks with inclinations
more face-on than $\cos{i}=0.84$ is $\sim 2-3$.
In order to reconcile the observed central line width of NGC~4143 with
the $M_{\rm BH}-\sigma_{\star}$ relation within the Keplerian-disk
framework, the disk would need to have an inclination angle of $27^{\circ}$.
However, it should be noticed that NGC~4143 is among our sample one of
the galaxies with a nearly unresolved spatial emissivity distribution
(see Fig.  \ref{fig:UppLim_FluxProfilesAndFitAndContinuum}), which
could actually be more concentrated than the one adopted by us.
Since this would lower the upper limit on $M_{\rm BH}$, we think that
NGC~4143 represents an interesting candidate for future
investigations.

The situation is different for the three galaxies with the lowest values of 
$\sigma_{\star}$ in our sample because their spatial flux
profiles are resolved.
Hence, we cannot explain their relatively high values of
$\sigma_{cen}$ ($\sim 100$ \kms) within the context of a Keplerian
disk in terms of gas orbiting in the vicinity of a $\sim 2\times10^6$
M$_{\odot}$ SMBH (as predicted by $M_{\rm BH}-\sigma_{\star}$
relation).  
Furthermore, as the derived sensitivity limits on $M_{\rm BH}$ for
these three galaxies are also around $\sim 2\times10^6$ M$_{\odot}$
(see Table \ref{tab:UppLim_Results}), we cannot expect the stellar
mass contribution to help explaining their $\sigma_{cen}$ values.
Indeed, the line widths obtained from the stellar potential {\it
and\/} a $2\times10^6$ M$_{\odot}$ SMBH are considerably smaller than
the observed ones. The predicted line widths in this case are for
these three galaxies in average $\sim 40$ and $\sim 75$
\kms, in the nearly face-on ($\cos{i}=0.84$) and edge-on ($\cos{i}=0.16$)
case, respectively.
Alternatively, the observed central line widths might arise in all
these three objects from highly inclined nuclear disks.  This is not
only unlikely but may also be insufficient, as in the case of
NGC~3982, even when considering a perfectly edge-on nuclear disk.
Therefore, in the case of the three less massive bulges, we may have
indirect evidence that at least part of the observed line width is due to
non-gravitational effects.

%
%
%
\section{Discussion and Conclusions}
\label{sec:UppLim_Disc&Concl}

We have demonstrated that with {\it HST}'s spatial resolution the integrated
line widths of the central emission lines provide stringent and
interesting constraints on the presence of SMBHs.  The relative
observational ease of this approach makes it potentially applicable to
large galaxy samples, that would allow us to test the universal applicability
of the emerging relations between $M_{\rm BH}$ and galaxy properties.

Our modeling, which was necessary to connect the observed
$\sigma_{cen}$ with the quantity of immediate interest,
$v_c(R_{ref})$, was based on the assumption that the gas line width
arises solely from orbital motion within a randomly oriented disk
around a putative SMBH.

Reality is undoubtedly more complex, and we have considered other potentially 
relevant effects, such as the stellar contribution to the total gravitational 
potential, and more simplistically, hydrostatic support of the gas. The 
dynamical influence of outflows and magnetic fields could also be important.
Except for fine-tuned circumstances, all these effects will provide an
additional contribution to the observed line width and the inferred
upper limit on $M_{\rm BH}$ will be tighter. 
Hence, our adopted set of assumptions lead to conservative estimates.

Comparison of our upper limits with direct $M_{\rm BH}$ determinations,
 either statistically (Fig.
\ref{fig:UppLim_UppLimKeplerian_G00_Mbhs}) or in a few cases individually
(Fig. \ref{fig:UppLim_UppLimKeplerian}), showed that our 1-$\sigma$ upper 
limits are generally near the actual value of $M_{\rm BH}$.

We have applied this analysis to a set of 16 galaxies whose sample
selection was not biased toward particular $M_{\rm BH}$ values.
Remarkably, with one exception, our $+1\sigma$ upper limits on $M_{\rm
BH}$ closely parallel the $M_{\rm BH}-\sigma_{\star}$ relation and
suggest that for galaxies with $\sigma_{\star}\geq 100\,{\rm
km\,s^{-1}}$, SMBHs with exceptionally high $M_{\rm BH}$ that violate
the $M_{\rm BH}-\sigma_{\star}$ relation must be rare.
By considerably broadening the range of host galaxies surveyed for
SMBHs, our 16 upper limits further support the emerging picture
wherein the black-hole mass and the overall galaxy structure are
closely linked.

Even with a limited sample of 16 objects, we have been able to isolate
a few cases worthy of further investigations.  NGC~4143 stands out as
the only object that falls below the $M_{\rm BH}-\sigma_{\star}$
relation; we speculate that this may indicate that its nuclear disk is
nearly face-on.  Three low-$\sigma_{\star}$ galaxies (NGC 3982, NGC
4321, and NGC 4548) seem to have $M_{\rm BH}$ upper limits that lie
systematically offset from other galaxies of low velocity dispersion
in which the $M_{\rm BH}$ was obtained by studying the stellar
kinematics. This suggests that in low-mass bulges non-gravitational
forces can considerably affect the gas motions in the central 10 pc.

\acknowledgments{This research was supported financially through NASA 
grant NAG 5-3556, and by GO-07361-96A, awarded by STScI, which is
operated by AURA, Inc., for NASA under contract NAS5-26555. Research
by A. J. Barth is supported by a postdoctoral fellowship from the
Harvard-Smithsonian Center for Astrophysics. A. V. Filippenko thanks
the Guggenheim Foundation for a Fellowship. M. Sarzi gratefully
acknowledges W. Dehnen and J. Magorrian for valuable comments and
suggestions on this work.}

%
%
%

%
%

\begin{deluxetable}{cccccccccc}
\footnotesize
\tablecaption{Basic Parameters of the Sample Galaxies \label{tab:UppLim_GalSample}}
\tablewidth{0pt}
\tablehead{
\colhead{Galaxy} & \colhead{Hubble Type} & \colhead{$B_T$} & \colhead{Spectral Class} & 
\colhead{$D$} & \colhead{$\sigma_{\star}$} & \colhead{Ref.} & \colhead{$\sigma_{e}$} & \colhead{$\sigma_{cen}$} & \colhead{Obs. Date} 
\\
\colhead{ } & \colhead{ } & \colhead{(mag)} & \colhead{} & 
\colhead{Mpc} & \colhead{(km s$^{-1}$)} & \colhead{} & \colhead{(km s$^{-1}$)} & \colhead{(km s$^{-1}$)} & \colhead{ }  
\\
\colhead{(1)} & \colhead{(2)} & \colhead{(3)} & \colhead{(4)} & 
\colhead{(5)} & \colhead{(6)} & \colhead{(7)} & \colhead{(8)} & \colhead{(9)} & \colhead{(10)} 
}
\startdata
NGC~2787 & SB0$^+$        & 11.82  & L1.9  & 13.0 & $210 \pm 23$ & 1  & $185 \pm 20$ & $215.1 \pm 4.3$ & 05 Dec. 1998 \\
NGC~3351 & SBb            & 10.53  & H     &  8.1 & $101 \pm 16$ & 2  & $ 93 \pm 15$ & $ 47.2 \pm 1.5$ & 25 Dec. 1998 \\
NGC~3368 & SABab          & 10.11  & L2    &  8.1 & $135 \pm 10$ & 3  & $114 \pm  8$ & $101.5 \pm 3.1$ & 31 Oct. 1998 \\
NGC~3982 & SABb:          & \ldots & S1.9  & 17.0 & $ 78 \pm  2$ & 4  & $ 78 \pm  2$ & $136.7 \pm 3.5$ & 11 Apr. 1998 \\
NGC~3992 & SBbc           & 10.60  & T2:   & 17.0 & $140 \pm 20$ & 4  & $119 \pm 17$ & $109.5 \pm 3.1$ & 19 Feb. 1999 \\
NGC~4143 & SAB0$^{\circ}$ & 11.65  & L1.9  & 17.0 & $270 \pm 12$ & 5  & $271 \pm 12$ & $226.3 \pm 2.1$ & 20 Mar. 1999 \\
NGC~4203 & SAB0$^-$:      & 11.80  & L1.9  &  9.7 & $124 \pm 16$ & 1  & $110 \pm 14$ & $148.9 \pm 5.2$ & 18 Apr. 1999 \\
NGC~4321 & SABbc          & 10.05  & T2    & 16.8 & $ 83 \pm 12$ & 6  & $ 74 \pm 11$ & $ 85.7 \pm 1.7$ & 23 Apr. 1999 \\
NGC~4450 & Sab            & 10.90  & L1.9  & 16.8 & $130 \pm 17$ & 2  & $121 \pm 16$ & $162.4 \pm 1.7$ & 31 Jan. 1999 \\
NGC~4459 & S0$^+$         & 11.32  & T2:   & 16.8 & $189 \pm 21$ & 1  & $167 \pm 18$ & $193.1 \pm 5.2$ & 23 Apr. 1999 \\
NGC~4477 & SB0:?          & 11.38  & S2    & 16.8 & $156 \pm 12$ & 7  & $134 \pm 10$ & $128.9 \pm 2.2$ & 23 Apr. 1999 \\
NGC~4501 & Sb             & 10.36  & S2    & 16.8 & $151 \pm 17$ & 8  & $136 \pm 15$ & $110.8 \pm 1.8$ & 26 Apr. 1999 \\
NGC~4548 & SBb            & 10.96  & L2    & 16.8 & $ 82 \pm  9$ & 9  & $ 71 \pm  8$ & $ 81.2 \pm 1.8$ & 26 Apr. 1999 \\
NGC~4596 & SB0$^+$        & 11.35  & L2::  & 16.8 & $154 \pm  5$ & 10 & $136 \pm  4$ & $142.0 \pm 8.7$ & 20 Dec. 1998 \\
NGC~4698 & Sab            & 11.46  & S2    & 16.8 & $134 \pm  6$ & 9  & $116 \pm  5$ & $101.9 \pm 2.2$ & 24 Nov. 1997 \\
NGC~4800 & Sb             & 12.30  & H     & 15.2 & $112 \pm  2$ & 4  & $112 \pm  2$ & $ 71.9 \pm 7.0$ & 03 Mar. 1999 \\

\enddata
\tablecomments{
Col. (1): Galaxy name. Col. (2): Hubble type from de~Vaucouleurs et
al. (1991).  Col. (3): Total apparent $B$ magnitude from
de~Vaucouleurs et al. (1991).  Col. (4): Nuclear spectral class from
Ho et al. (1997), where H = \ion{H}{2} nucleus, L = LINER, S =
Seyfert, T = ``transition object'' (LINER/\ion{H}{2}), 1 = type~1, 2 =
type~2, and a fractional number between 1 and 2 denotes various
intermediate types; uncertain and highly uncertain classifications are
followed by a single and double colon, respectively.  Col. (5):
Distance from Tully (1988), who assumes $H_0$ = 75 \kms\ Mpc$^{-1}$.
Col. (6): Ground-based central stellar velocity dispersion
$\sigma_{\star}$.  Col. (7): Reference for $\sigma_{\star}$.  Col. (8)
Stellar velocity dispersion within one effective radius $\sigma_e$,
derived from the $\sigma_{\star}$ value following the extrapolation of
J{\o}rgensen, Franx, \& Kjaergaard (1995).  Col. (9): Measured gas
velocity dispersion within the central $0\farcs25 \times 0\farcs2$.
Col. (10): UT observation date.\\ References. --- (1) Dalle Ore \etal
1991; (2) Whitmore, Schechter, \& Kirshner 1979; (3) H{\'e}raudeau et
al. 1999; (4) this paper (\S
\ref{subsec:UppLim_Observations}); (5) Di~Nella \etal 1995; (6)
Schechter 1983; (7) Jarvis \etal 1988; (8) H{\'e}raudeau \& Simien
1998; (9) Corsini \etal 1999; (10) Kent 1990.
}
\end{deluxetable}

\begin{deluxetable}{cccccc}
\footnotesize
\tablecaption{Results of the Rotating Disk Model \label{tab:UppLim_Results}}
\tablewidth{0pt}
\tablehead{
\colhead{Galaxy Name} & \colhead{$\sigma_{flux}$} & \colhead{$\sigma_{flux,corr}$} & 
\colhead{$M_{\rm BH}(\Phi=\Phi_{\rm BH})$} & \colhead{$M_{\rm BH}(\Phi=\Phi_{\rm BH}+ \Phi_{\star})$} & 
\colhead{$M_{\rm BH}$(Sensitivity Limit)}  
\\
\colhead{ } & \colhead{(arcsec)} & \colhead{(arcsec)} & 
\colhead{ (M$_{\odot}$)} & \colhead{(M$_{\odot}$)} & \colhead{(M$_{\odot}$)}   
\\
\colhead{(1)} & \colhead{(2)} & \colhead{(3)} & \colhead{(4)} & 
\colhead{(5)} & \colhead{(6)} 
}
\startdata
NGC~2787 & 0.056 & 0.053 & 1.9$\times10^8$ & 1.8$\times10^8$ & 3.8$\times10^6$ \\
NGC~3351 & 0.132 & 0.153 & 9.7$\times10^6$ & 8.0$\times10^6$ & 2.1$\times10^6$ \\
NGC~3368 & 0.086 & 0.095 & 3.8$\times10^7$ & 2.7$\times10^7$ & 4.5$\times10^6$ \\
NGC~3982 & 0.051 & 0.046 & 8.0$\times10^7$ & 7.5$\times10^7$ & 3.3$\times10^6$ \\
NGC~3992 & 0.051 & 0.046 & 5.7$\times10^7$ & 5.3$\times10^7$ & 3.0$\times10^6$ \\
NGC~4143 & 0.035 & 0.024 & 1.4$\times10^8$ & 1.4$\times10^8$ & 1.4$\times10^6$ \\
NGC~4203 & 0.030 & 0.015 & 2.3$\times10^7$ & 2.3$\times10^7$ & 4.3$\times10^5$ \\
NGC~4321 & 0.046 & 0.040 & 2.7$\times10^7$ & 2.5$\times10^7$ & 2.3$\times10^6$ \\
NGC~4450 & 0.046 & 0.040 & 1.1$\times10^8$ & 1.1$\times10^8$ & 3.1$\times10^6$ \\
NGC~4459 & 0.041 & 0.033 & 1.3$\times10^8$ & 1.3$\times10^8$ & 2.9$\times10^6$ \\
NGC~4477 & 0.056 & 0.053 & 8.7$\times10^7$ & 7.8$\times10^7$ & 5.1$\times10^6$ \\
NGC~4501 & 0.081 & 0.088 & 9.0$\times10^7$ & 7.4$\times10^7$ & 7.6$\times10^6$ \\
NGC~4548 & 0.066 & 0.067 & 3.3$\times10^7$ & 2.8$\times10^7$ & 3.8$\times10^6$ \\
NGC~4596 & 0.056 & 0.053 & 1.1$\times10^8$ & 9.4$\times10^7$ & 5.6$\times10^6$ \\
NGC~4698 & 0.091 & 0.100 & 8.0$\times10^7$ & 7.1$\times10^7$ & 6.0$\times10^6$ \\
NGC~4800 & 0.076 & 0.081 & 3.2$\times10^7$ & 2.0$\times10^7$ & 6.1$\times10^6$ \\

\enddata
\tablecomments{
Col. (1): Galaxy name.  Col. (2): Derived first guess of the intrinsic
flux extent(see \S \ref{subsec:UppLim_TheDiskModelling}).  Col. (3):
Adopted intrinsic flux extent(see \S
\ref{subsec:UppLim_TheDiskModelling}).  Col (4): Upper limit on
$M_{\rm BH}$ ($+1\sigma$) obtained with a rotating disk model ignoring
the stellar gravitational potential (\S
\ref{subsec:UppLim_TheDiskModelling}).  Col (5): Upper limit on
$M_{\rm BH}$ ($+1\sigma$) obtained with a rotating disk model
including the stellar gravitational potential (\S
\ref{subsec:UppLim_TheDiskModelling}).  Col. (6) Sensitivity limit to
$M_{\rm BH}$.
}
\end{deluxetable}

\end{document}